\pdfoutput=1 
\documentclass[superscriptaddress,longbibliography,aps,prl,floatfix,reprint]{revtex4-1}
\usepackage{float}
\usepackage{bbold}
\usepackage{graphicx}
\graphicspath{{figs/}}
\usepackage[caption=false,position=top]{subfig}
\usepackage{hyperref}
\usepackage{xcolor}
\usepackage{physics}
\usepackage[normalem]{ulem}
\hypersetup{colorlinks=true, citecolor=blue, urlcolor=blue, linkcolor=blue}

\newcommand{\balpha}{\boldsymbol{\alpha}}
\newcommand{\bbeta}{\boldsymbol{\beta}}

\makeatletter
\pdfpageheight\paperheight
\pdfpagewidth\paperwidth
\makeatother

\begin{document}
\title{Making Trotters Sprint: A Variational Imaginary Time Ansatz for Quantum Many-body Systems}

\author{Matthew J.~S.~Beach}
\affiliation{Perimeter Institute for Theoretical Physics, Waterloo, Ontario N2L 2Y5, Canada}
\affiliation{Department of Physics and Astronomy, University of Waterloo, Waterloo N2L 3G1, Canada}
\author{Roger G.~Melko}
\affiliation{Perimeter Institute for Theoretical Physics, Waterloo, Ontario N2L 2Y5, Canada}
\affiliation{Department of Physics and Astronomy, University of Waterloo, Waterloo N2L 3G1, Canada}
\author{Tarun Grover}
\affiliation{Department of Physics, University of California at San Diego, La Jolla, CA 92093, USA}
\author{Timothy H.~Hsieh}
\affiliation{Perimeter Institute for Theoretical Physics, Waterloo, Ontario N2L 2Y5, Canada}
\date{\today}

\begin{abstract}
    We introduce a variational wavefunction for many-body ground states that involves imaginary time evolution with two different Hamiltonians in an alternating fashion with variable time intervals.
    We successfully apply the ansatz on the one- and two-dimensional transverse-field Ising model and
    systematically study its scaling for the one-dimensional model at criticality.
    We find the total imaginary time required scales logarithmically with system size, in contrast to the linear scaling in conventional Quantum Monte Carlo.
    We suggest this is due to unique dynamics permitted by alternating imaginary time evolution, including exponential growth of bipartite entanglement.
    For generic models, the superior scaling of our ansatz potentially mitigates the negative sign problem at the expense of having to optimize variational parameters.
\end{abstract}
\maketitle

\emph{Introduction--}
Imaginary time plays a prominent role in multiple branches of physics, including cosmology, statistical mechanics and quantum field theory.
The seemingly simple replacement of real time, $t$, with its imaginary counterpart, $\tau = -it$, leads to fundamental connections between quantum theory and statistical mechanics
\cite{wickPropertiesBetheSalpeterWave1954}.
Such connections enable the efficient simulation of many quantum systems using quantum Monte Carlo techniques
\cite{handscombMonteCarloMethod1962,
    blankenbeclerProjectorMonteCarlo1983,
    sandvikQuantumMonteCarlo1991}.
However, for many physically interesting models, these methods suffer from the prohibitive `negative sign problem'
\cite{lohSignProblemNumerical1990,
    troyerComputationalComplexityFundamental2005},
which requires an exponential amount of computational resources to obtain reasonable accuracy for quantum many-body systems.
Many outstanding problems in condensed matter, such as those involving high temperature superconductors or topologically ordered phases, require an understanding of complex interacting models which are unsolved with present techniques.

One class of Monte Carlo methods that can avoid the sign problem are so-called variational Monte Carlo (VMC) methods
\cite{gutzwillerEffectCorrelationFerromagnetism1963,
    *gutzwillerCorrelationElectronsNarrow1965,
    ceperleyMonteCarloSimulation1977,
    otsukaVariationalMonteCarlo1992,
    baeriswylVariationalSchemesManyElectron1987,
    beccaQuantumMonteCarlo2017}.
In VMC, one assumes a sufficiently general trial state that depends on adjustable parameters. These parameters are then chosen to minimize the energy with respect to the given Hamiltonian.
Finding an effective trial state such as Jastrow
\cite{beccaQuantumMonteCarlo2017},
matrix product states
\cite{verstraeteDensityMatrixRenormalization2004,
    sandvikVariationalQuantumMonte2007,
    vidalClassicalSimulationInfiniteSize2007},
or neural network states
\cite{carleoSolvingQuantumManybody2017,
    freitasNeuralNetworkOperations2018,
    inackProjectiveQuantumMonte2018,
    pilatiSelflearningProjectiveQuantum2019},
can result in efficient simulation of interacting quantum systems.
The key to the success of these techniques is a well-chosen ansatz that reflects the properties of the target phase and the existence of a viable optimization scheme
\cite{
    umrigarOptimizedTrialWave1988,
    sorellaGreenFunctionMonte1998,
    *sorellaGeneralizedLanczosAlgorithm2001,
    *sorellaWaveFunctionOptimization2005}.

The recent advent of quantum computers and simulators has motivated the development of new variational approaches
\cite{jonesVariationalQuantumAlgorithms2019}.
Such variational quantum algorithms involve applying a sequence of unitary operators, parameterized by several variables onto a easy-to-prepare initial state.
The variables are chosen to optimize a given cost function involving the resulting wavefunction.
For example, in the quantum approximate optimization algorithm (QAOA)
\cite{farhiQuantumApproximateOptimization2014,
    zhouQuantumApproximateOptimization2018,
    wangQuantumApproximateOptimization2018,
    hadfieldQuantumApproximateOptimization2019,
    verdonQuantumApproximateOptimization2019}
the cost function is a classical Hamiltonian encoding a combinatorial optimization problem, and the variational wavefunction is prepared by alternating between evolving with the Hamiltonian and a transverse field.
The evolution times constitute variational parameters that are optimized to minimize the Hamiltonian cost function.
This variational approach has been generalized for preparing both strongly correlated
and highly-entangled states on near-term quantum devices
\cite{weckerProgressPracticalQuantum2015,
    hoEfficientVariationalSimulation2019,
    hoUltrafastVariationalSimulation2019}.

Motivated by the success of such variational approaches,
in this Letter we propose a variational ansatz for ground states of quantum many-body systems which involves sequentially evolving with different Hamiltonians in imaginary time.
In contrast to real time evolution with local Hamiltonians, which is limited by Lieb-Robinson bounds on the growth of correlation functions, imaginary time evolution does not have this constraint and can exhibit remarkable efficiency in traversing Hilbert space.

As proof of concept, we demonstrate the efficiency of our ansatz in representing the ground state of the transverse field Ising model at criticality.
Whereas standard projector methods require imaginary time scaling with system size $L$ to reach the critical ground state, we show numerically that our ansatz requires time scaling logarithmically with $L$.
Furthermore, we analyze how entanglement grows after each imaginary time operation in our ansatz, and we find an exponential growth that is a unique feature of imaginary time dynamics.
We conclude by demonstrating that the ansatz continues to perform well in the presence of integrability-breaking perturbations, and we mention generalizations of our ansatz to other models, including those with sign problems.
We envision the main purpose of this ansatz to be an efficient trial wavefunction for quantum many-body physics on (classical) computers, however, it is possible that one can also implement such imaginary time evolution natively on a quantum computer
\cite{mcardleVariationalQuantumSimulation2018,
    mottaQuantumImaginaryTime2019}.

\begin{figure}[thbp]
    \centering
    \includegraphics[trim={185 5 50 45},clip,width=0.85\columnwidth]{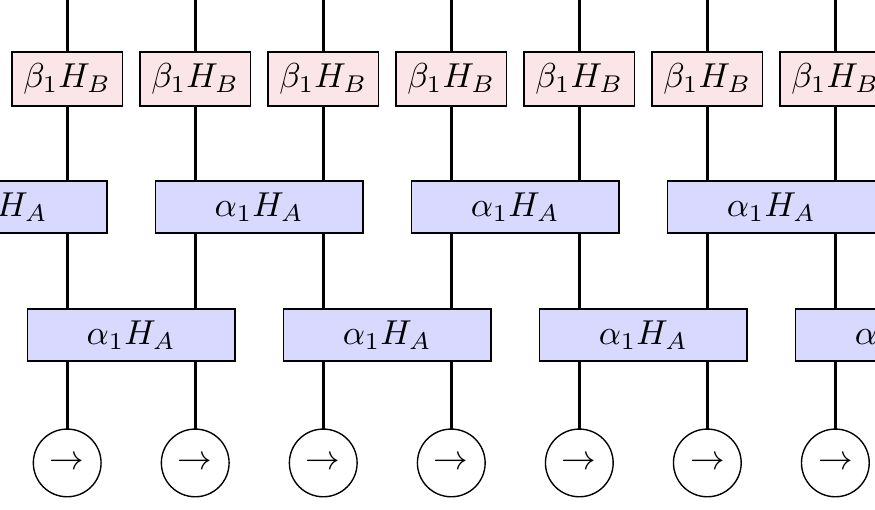}
    \caption{\label{fig:diagram}
        Circuit representation of the trial state $\ket{\psi_P(\balpha, \bbeta)}$ for $P=1$.  Each box denotes imaginary time evolution with the enclosed Hamiltonian.
    }
\end{figure}

\emph{Ansatz--}
Many Hamiltonians are naturally a linear combination of two components $H=H_A + g H_B$, where $H_{A,B}$ are individually tractable to analyze.
Examples include transverse field Ising, Hubbard, and the $J_1$--$J_2$ model.
Motivated by the QAOA procedure,
we consider the following variational imaginary time ansatz (VITA) for the ground state of $H$:
\begin{align}
    \label{eq:trial}
    \ket{\psi_P(\balpha,\bbeta)} & = \mathcal{N} e^{-\beta_P H_B} e^{-\alpha_P H_A} \cdots e^{-\beta_1 H_B} e^{-\alpha_1 H_A} \ket{\psi_0}
\end{align}
where $P$ is the number of pairs of variational parameters $\balpha=(\alpha_1,...,\alpha_P),\bbeta=(\beta_1,...\beta_P)$, $|\psi_0\rangle$ is an initial state, and $\mathcal{N}$ is a normalization factor. We further define the total imaginary time $\tau = \frac{1}{2}\sum_{p=1}^P (\alpha_p + \beta_p)$.
A circuit representation is shown in Fig.~\ref{fig:diagram}.
The bang-bang, or square-pulse, style of the ansatz is optimal for quantum control in real-time QAOA as per Pontryagin's principle \cite{yangOptimizingVariationalQuantum2017}.

While VITA is applicable to any Hamiltonian, in specific cases there is explicit physical motivation for considering such an ansatz.
For example, for the fermionic Hubbard model, the $P=1$ ansatz reduces to
Otsuka's generalization of the Gutzwiller variational wavefunction
\cite{otsukaVariationalMonteCarlo1992,
    gutzwillerEffectCorrelationFerromagnetism1963,
    *gutzwillerCorrelationElectronsNarrow1965,
    baeriswylVariationalSchemesManyElectron1987},
which seeks to balance single occupancy per site with itinerancy.
The $P \leq 3$ case has been considered in Ref.~\cite{yanagisawaOffDiagonalWaveFunction1998,
    *yanagisawaCrossoverWeaklyStrongly2016,
    *yanagisawaAntiferromagnetismSuperconductivityPhase2019}
for the two-dimensional Hubbard model, but a systematic analysis of how its performance scales with system size and $P$ was not carried out.
A related variational approach for the Hubbard model has also been considered in
Ref.~\cite{vaeziUnifiedTheoryVariational2018}.

The standard projector method for attaining the ground state of $H$ involves evaluating $e^{-\tau H}|\psi_0\rangle$ for $\tau \gtrsim 1/\Delta$ where $\Delta$ is the many-body spectral gap.
This can be decomposed via Trotterization into a sequence of the VITA form, with  parameters $\alpha_p = \beta_p = \tau/2P$ for large $P$.
This guarantees that VITA can exactly represent the ground state in the $P\rightarrow\infty$ limit.
However, the projector method is especially expensive for critical systems where $\Delta \sim 1/L$, and hence $\tau$ scales polynomially with $L$.
One can consider Eq.~(\ref{eq:trial}) as a non-uniform Trotterization with large (and variable) times steps.
We will show that remarkably high fidelities can be attained even with $\tau$ that is  \textit{exponentially smaller} compared to the aforementioned estimate from the standard projector method.

We first present some general considerations of why such an ansatz may be efficient.
It is useful to first compare with the real-time analogue, which are QAOA-type circuits involving alternating real-time evolution between two Hamiltonians.
The Lieb-Robinson bound dictates that real-time evolution with local Hamiltonians can generate correlations only within a light cone, and thus there are lower bounds on the time it takes to prepare highly correlated states starting from unentangled product states.
For example, in one dimension, the total time to prepare the GHZ (``cat'') state
$\frac{1}{\sqrt2}\left(|11\ldots 1 \rangle + |00\ldots 0\rangle\right)$
scales at least linearly with system size $L$
\cite{bravyiLiebRobinsonBoundsGeneration2006,
    hoEfficientVariationalSimulation2019}.
In contrast, by evolving with the GHZ parent Hamiltonian $H_{ZZ} = -\sum_{i=1}^L Z_i Z_{i+1}$ in imaginary time ($Z$ is the Pauli-$Z$ matrix), the GHZ state can be prepared with imaginary time scaling as $\log L$
\footnote{See Supplemental Material}.

\emph{Application--}
We test VITA on the transverse-field Ising model (TFIM)
\begin{align}\label{eq:ham}
    H & = H_{ZZ} + h H_{X}  \equiv -\sum_{i=1}^N Z_i Z_{i+1} - h \sum_{i=1}^N X_i
\end{align}
with periodic boundary conditions on a system with $N$ spins.
$Z, X$ are the Pauli matrices, and $h$ is the transverse field strength.
Our ansatz in this case starts from the paramagnetic ground state of $H_X$, $|+\rangle$ and alternates between $H_A=H_{ZZ}$ and $H_B = H_X$.

The Ising chain can be mapped to free fermions via the Jordan-Wigner transformation \cite{liebTwoSolubleModels1961} which allows for the efficient evaluation of Eq.~(\ref{eq:trial}).
We can thus optimize our ansatz for very large system sizes and several pulses.
This allows us to properly characterize how efficient the VITA ansatz is without introducing sampling error.
For a fixed $P$, optimization involves finding the minima of the energy cost function
$E_P(\balpha, \bbeta) = \ev{H}{\psi_P(\balpha,\bbeta)}$.

\begin{figure}[tbhp]
    \includegraphics[width=\columnwidth]{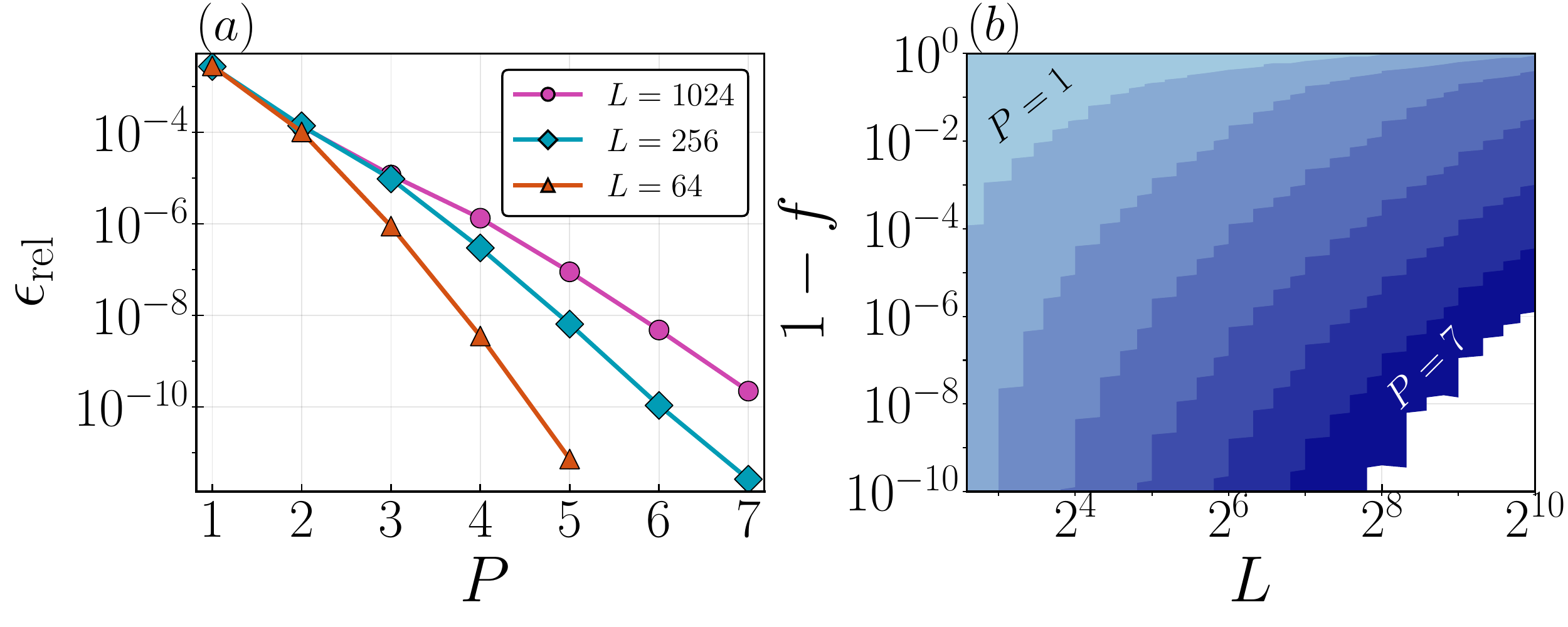}
    \caption{
        \label{fig:energy}
        (a) Relative error in energy, $\epsilon_{\rm rel}$
        between the exact ground state energy, $E_{\rm exact}$,
        and the energy of the optimized trial wavefunction $E_P(\balpha, \bbeta)$.
        (b) Number of pulses $P$ needed to obtain a desired accuracy in the fidelity, $f$, for a given system size, $L$.
        The white region was not computed in the present study.
    }
\end{figure}

We first focus on approximating the critical ground state of $H$ at $h=1$.
Figure~\ref{fig:energy}a shows the relative error in energy
$\epsilon_{\rm rel} = |(E_P(\balpha, \bbeta) - E_{\rm exact})/E_{\rm exact}|$
where $E_{\rm exact}$ is the exact ground state energy at the critical point, for various $P$.
Evidently, increasing $P$ dramatically improves the accuracy in the energy, even for large system sizes.  Optimized parameters for various $P$ are provided in supplementary material.

Since the exact ground state for the TFIM is known, we also compare the fidelity, $f$, of the optimized trial state with the target state.
The error in fidelity,
$1-f \equiv 1-|\langle \psi_{\rm exact}|\psi_P(\balpha, \bbeta)\rangle\rangle|^2$,
is shown in Fig.~\ref{fig:energy}b for various $P, L$.
The efficiency is quite remarkable; for example, for $L=64$, $P=2$ is already sufficient to approximate the critical state to within around $10^{-4}$ in relative energy and $10^{-2}$ in fidelity.
Recall that the error in fidelity provides an upper-bound for the error in any observable
\footnote{See Supplemental Material}.

\emph{Entanglement dynamics--}
While there is no Lieb-Robinson bound limiting the rate for generating long-range correlations in our ansatz, entanglement considerations provide lower bounds on the iterations $P$ required to prepare the critical state.
Ignoring normalization, imaginary time evolution with a local Hamiltonian can be represented by a (non-unitary) quantum circuit; each iteration of our ansatz corresponds to three layers shown in Fig.~1.
After $P$ pulses, the bipartite entanglement entropy between the left and right halves the system, hereafter abbreviated EE, can be attained by bisecting the circuit through $P$ bonds.
Hence, EE after $P$ iterations is at most $P \log{D}$, where $D$ is the Schmidt rank (number of singular values) upon decomposing a single two-qubit imaginary time operator.

In order to generate the EE of the critical state,
which scales as $S \propto \log{L}$,
we need the number of pulses scaling at least as $P \propto \log{L}$.
We observe that Fig.~\ref{fig:energy}b is consistent with this scaling form.
For example, for a target fidelity error of $10^{-10}$, each additional $P$ in the ansatz can represent a system approximately twice as large.

\begin{figure}[bhtp]
    \centering
    \includegraphics[width=\columnwidth]{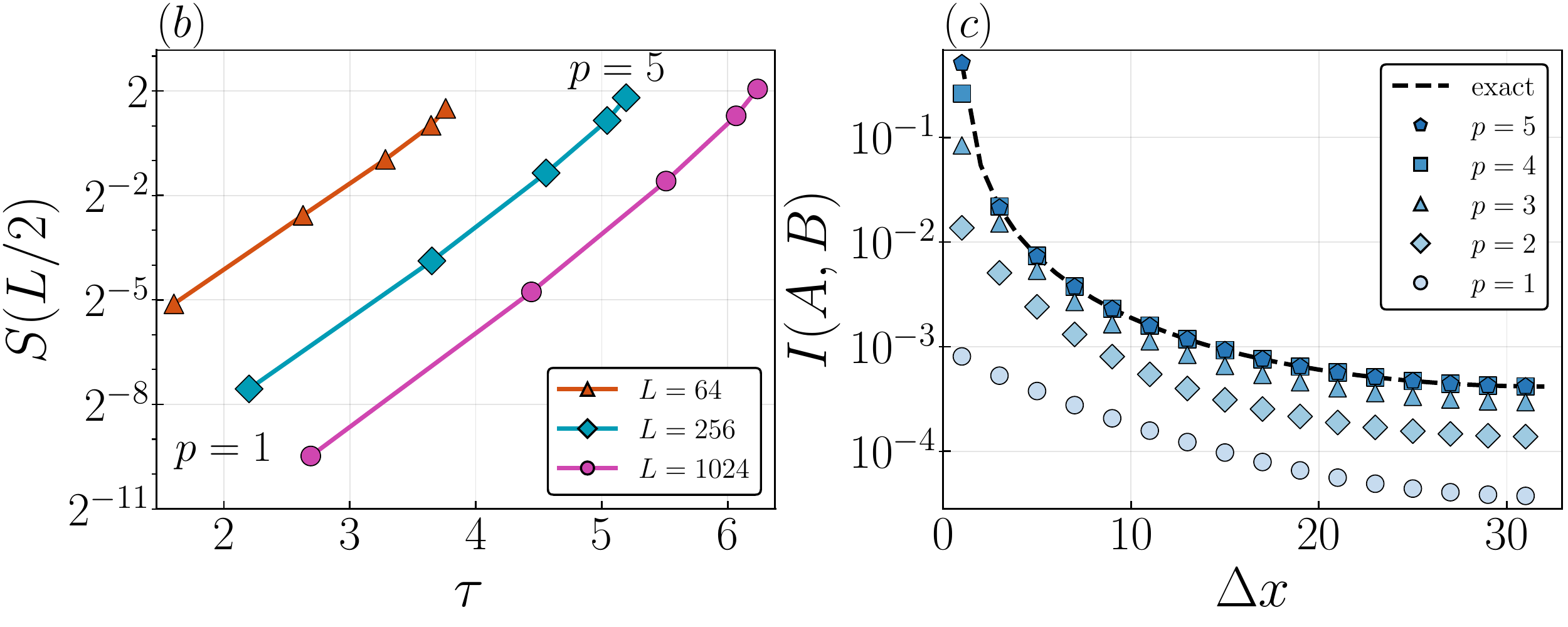}
    \caption{\label{fig:entropy}
        (a) Entanglement entropy of half partition grows exponentially with imaginary time $\tau$, in the optimal
        $P=5$ ansatz for the critical state.
        (b) Mutual information between two spins at positions $A$ and $B$ respectively as a function of their distance $\Delta x$, for intermediate steps $p$ in the $P=5$ protocol with $L=64$.
    }
\end{figure}

The entanglement dynamics in imaginary time evolution can be considerably different from its real time counterpart.
For real time evolution, EE across a bipartition can increase only by acting with an operator supported on both sides of the partition.
If the circuit in Fig.~\ref{fig:diagram} were unitary, any increase in EE from one layer to the next would be bounded by a constant depending on the two-qubit unitary but not on the state being acted on (see the ``small incremental entangling theorem'' \cite{marienEntanglementRatesStability2014}).
In contrast, even imaginary time operators acting on one side of the bipartition can generate entanglement across the cut.
As a very simple example involving two spins, the action of
$e^{-\beta Z_1}$ on
$ \eta_{+} |11\rangle + \eta_{-} |00\rangle$
can increase EE as long as $|\eta_{+}|>|\eta_{-}|>0$.
This illustrates that the more entangled the initial state, the more imaginary time evolution can change the entanglement.
The change is not simply bounded by a state-independent constant.
This allows in principle an exponential growth of EE, as long as the total EE after $P$ steps lies below $P \log{D}$.

Our ansatz exhibits such dynamics.
We take the $P=5$ ansatz and analyze the EE of the states at intermediate steps, $p$, of our protocol using the technique of
\cite{peschelCalculationReducedDensity2003,
vidalEntanglementQuantumCritical2003,
*latorreGroundStateEntanglement2004}.
We find that the EE increases exponentially with imaginary time (Fig.~\ref{fig:entropy}a).
Moreover, for every intermediate state, we plot the mutual information ($S_A+S_B-S_{AB}$) between two spins $A, B$ as a function of their separation (Fig.~\ref{fig:entropy}b).
The power law decay for every step is in stark contrast to any local real time evolution and illustrates the ability of imaginary time evolution to generate long-range correlations
\footnote{See Supplemental Material at [URL will be inserted by publisher] for
    considerations of a local quench in imaginary time in a conformal field theory }.
We find that under imaginary-time evolution, entanglement starts to grow immediately  following a local quench, in contrast to real-time evolution
\cite{calabreseEvolutionEntanglementEntropy2005,
    *calabreseEntanglementCorrelationFunctions2007}
where it takes a time proportional to $\ell$ before growing,  $\ell$ being the distance between the location of the local quench and the entanglement cut.

\emph{Scaling--} To compare directly with the projector method,
we also investigate the total imaginary time $\tau \equiv \frac{1}{2}\sum_{p=1}^P (\alpha_p + \beta_p)$ required to achieve a target fidelity.
Motivated by the $P \propto \log{L}$ scaling for achieving a target fidelity, we propose a scaling form of
$1-f = G(\tau ( \log L)^{-\nu})$ for some exponent $\nu$.
We perform a scaling collapse for $L\in[4, 6, \ldots, 1024]$, and $P\in[1, \ldots, 7]$ and find the optimal exponent $\nu =  2.3 \pm 0.1$.
This logarithmic scaling is an exponential speedup compared to the linear scaling of the projector method,
$1-f = F(\tau L^{-1})$ \cite{liuTypicalityQuantumcriticalPoints2018}.

\begin{figure}[tbhp]
    \centering
    \includegraphics[width=0.72\columnwidth]{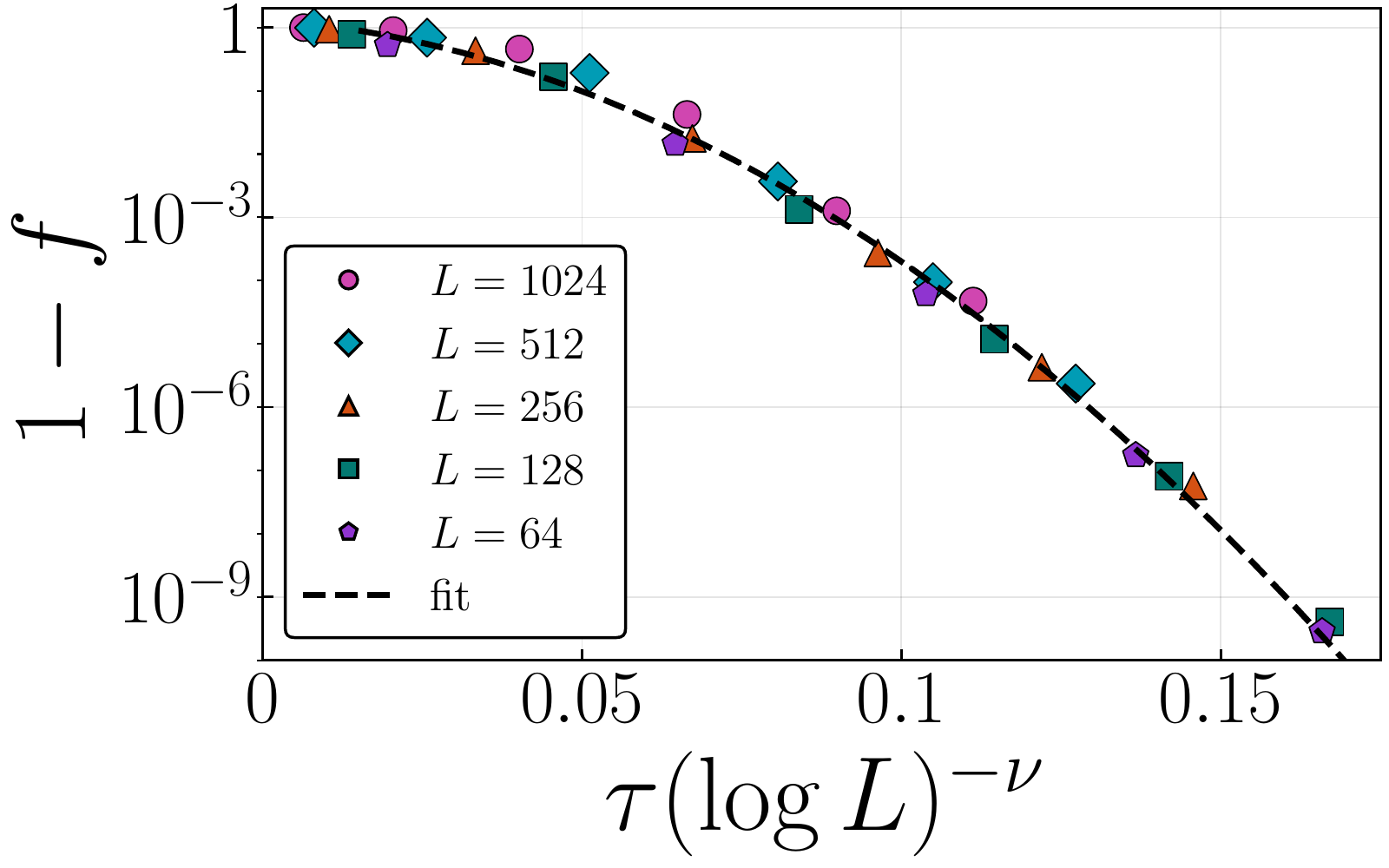}
    \caption{\label{fig:collapse}
        Collapse of the infidelity $\log(1-f)  = G(\frac{\tau}{(\log L) ^{\nu}})$ with $\nu = 2.3$. The fit is a power law $\log(1-f) = 175 \, x^{1.85}$ with $x = \frac{\tau}{(\log L) ^{\nu}}$.}
\end{figure}

\emph{Monte Carlo approach--} While the TFIM model admits a dual representation as free fermions,
for a general model
sampling methods are crucial for estimating the energy cost function.
As a proof of concept, we also use Monte Carlo sampling for stochastically optimizing VITA.

The quantum-classical correspondence maps quantum observables to dual classical observables of a classical anisotropic Ising model on an $L \times(2P+1) $ lattice.
We denote the classical spin configurations by $\{s\}$ and the spatial and imaginary time by $(i, p)$, respectively \footnote{
    This is a slight abuse of notation since $1\leq p \leq P$ in Eq.~(\ref{eq:trial}). However, this may be permitted since there are only $p$ unique time couplings $J_t(p)$ due to the lattice symmetry.}.

The expectation value of a quantum observable $\mathcal{O}$ is
\begin{align}
    \ev{\mathcal O}{\psi_P(\balpha,\bbeta)}
    = \sum_{\{s\}} \tilde{\mathcal{O}}(s) \, p_{\balpha,\bbeta}(s)
\end{align}
where
$\tilde{\mathcal{O}}$ are dual classical observables, and
$p_{\balpha,\bbeta}(s)$ is the Boltzmann weight corresponding to the Ising model with couplings
$J_x(p) = \alpha_p,  J_{\tau}(p) = \frac{1}{2} \ln \coth \beta_p$
between nearest neighbors in space and imaginary time, respectively.
The couplings vary with imaginary time, $p$, but are uniform in space.
In this way, the energy of the trial wavefunction can be sampled efficiently with Monte Carlo.

We use this scheme with $P=1,2$ to target the ground states for various values of the transverse field $h$ in both the (integrable) one-dimensional and the (non-integrable) two-dimensional TFIM.
We use stochastic natural gradient descent (stochastic reconfiguration) to optimize the parameters
\cite{sorellaGreenFunctionMonte1998}.
We find rapid convergence for $P=1$, while higher-$P$ becomes more difficult, especially for with a noisy objective function.

\begin{figure}[tbhp]
    \centering
    \includegraphics[width=\columnwidth]{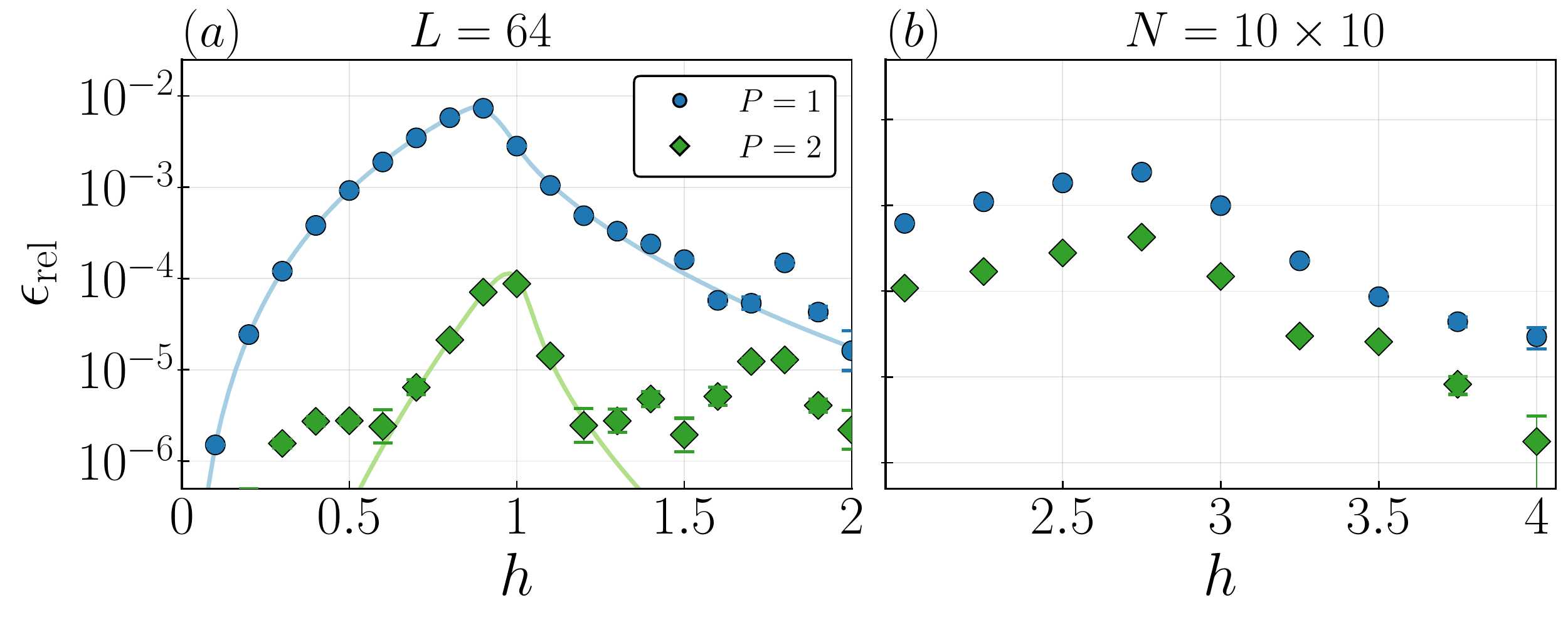}
    \caption{\label{fig:vmc}
        Relative error in energy, $\epsilon_{\rm rel}$ for VMC using our ansatz on the TFIM:
        (a) 1d model with $L=64$ spins. Solid lines denotes the results from the free fermion approach,
        (b) 2d model on a $10 \times 10$ square lattice. Energies are compared with those from zero-temperature stochastic series expansion
        \cite{sandvikGroundStateProjection2005,*sandvikStochasticSeriesExpansion2003}.
    }
\end{figure}

The relative error in energy achieved is shown in Fig.~\ref{fig:vmc},
with the free fermion results for comparison.
For $P=1$ the VMC achieves the same accuracy as the free fermion method.
However, for $P=2$ away from the critical point $h=1$, the VMC performance is limited by sampling error
\footnote{
    Of course the statistical error in Monte Carlo can be made arbitrarily small given long enough run-time since the
    sampling error goes as $\mathcal O (N_{\rm MC}^{-1/2})$ for $N_{\rm MC}$ Monte Carlo sweeps.
    This guarantees that VMC will converge to the free fermion solution in Fig.~\ref{fig:vmc}a if the globally optimal parameters can be found.
    The computational time of the MC sampling is $\mathcal O (\tau_{\rm corr} N_{\rm MC})$ where the autocorrelation time $\tau_{\rm corr}\sim (PN)^{\gamma}$ is a polynomial function of $PN$ for $N$ spins and $P$ pulses of VITA.
}. 

\emph{Discussion--} We have introduced a variational technique that is motivated by both projector methods and recently developed quantum algorithms.  It provides substantial shortcuts to the usual Trotterization of imaginary time evolution, at the expense of making the procedure variational.
Using TFIM as a first testbed, we have demonstrated that this ansatz is viable for sampling methods and highly efficient.
In particular, the number of variational parameters required to represent the TFIM critical state scales as $\sim \log{L}$, in contrast to other variational methods such as density matrix renormalization group (DMRG), which in this critical case requires bond dimension scaling with $L$ and thus number of parameters scaling with $L^2$.
One reason for this efficiency is the fact that imaginary time evolution is not subject to many bounds for real time evolution;
despite being generated by local Hamiltonians, our ansatz exhibits an exponential growth of entanglement entropy and rapid generation of long-range correlations, features unique to imaginary time evolution.

Our variational approach is potentially useful in the many situations where imaginary time Trotterization involves a prohibitively large number of steps.
For example, in models with a sign problem, the computational cost scales exponentially with space and imaginary time $\mathcal O (\tau L^d)$.
Our ansatz provides a variational shortcut that significantly reduces $\tau$ (from $\tau \sim L$ to $\tau\sim (\log L)^{2.3}$ in the critical 1d TFIM) which could enable the study of larger systems even with a sign problem.
For few variational parameters ($P=1,2$), the optimization may be feasible, and we leave these investigations to future work.

\begin{acknowledgments}
    \emph{Acknowledgments--}
    The authors would like to thank
    J.~Carrasquilla,
    L.~E.~Hayward Sierens,
    E.~Inack,
    B.~Kulchytskyy,
    for many useful discussions.
    This research was supported by the Natural Sciences and Engineering Research Council of Canada (NSERC), the Canada Research Chair program, and the Perimeter Institute for Theoretical Physics.
    This work was made possible by the facilities of the Shared Hierarchical Academic Research Computing Network (SHARCNET) and Compute/Calcul Canada.
    Research at Perimeter Institute is supported by the Government of Canada through Industry Canada and by the Province of Ontario through the Ministry of Research \& Innovation.
    TG is supported by the National Science Foundation under Grant No. DMR-1752417, and as an Alfred P. Sloan Research Fellow.
    This research was supported in part by the National Science Foundation under Grant No. NSF PHY-1748958.
\end{acknowledgments}

\bibliography{paper}

\end{document}

\title{Supplementary Material for: ``Making Trotters Sprint: A Variational Imaginary Time Ansatz for Quantum Many-Body Systems''}

\author{Matthew J.~S.~Beach}
\affiliation{Perimeter Institute for Theoretical Physics, Waterloo, Ontario N2L 2Y5, Canada}
\affiliation{Department of Physics and Astronomy, University of Waterloo, Waterloo N2L 3G1, Canada}
\author{Roger G.~Melko}
\affiliation{Perimeter Institute for Theoretical Physics, Waterloo, Ontario N2L 2Y5, Canada}
\affiliation{Department of Physics and Astronomy, University of Waterloo, Waterloo N2L 3G1, Canada}
\author{Tarun Grover}
\affiliation{Department of Physics, University of California at San Diego, La Jolla, CA 92093, USA}
\author{Timothy H.~Hsieh}
\affiliation{Perimeter Institute for Theoretical Physics, Waterloo, Ontario N2L 2Y5, Canada}
\date{\today}

\maketitle

\section{Jordan-Wigner Transformation}\label{app:JW}

The 1d TFIM conveniently admits a solution via the Jordan-Wigner transformation
\cite{liebTwoSolubleModels1961},
where $L$ Ising spins are mapped to $L/2$ independent spin-$\frac{1}{2}$ fermions by introducing the fermionic operators
\begin{align}
    a_j & = e^{i\pi \phi_j} S^-_j, \qquad
    a_j^{\dag}  = e^{-i\pi \phi_j} S^+_j, \qquad
\end{align}
with $\phi_j = \sum_{i=1}^{j-1} S^+_i S^-_i$ and where $S^{\pm}_j = (Y_j \pm i Z_j)/2$ are the raising/lowering operators.
Performing a Fourier transform, we have
\begin{align*}
    H_{ZZ} & = \sum_{k=0}^{L-1} \left( 2 c^\dag_k c_k -1 \right)
    \\
    H_{X}  & = 2\sum_{k=0}^{(L-1)/2} \cos \theta_k \left( c^{\dag}_k c_k + c^{\dag}_{-k} c_{-k}\right)
    \\
           & \qquad \qquad
    + i \sin \theta_k \left( c^{\dag}_k c_k + c^{\dag}_{-k} c_{-k}\right)
\end{align*}
with $\theta_k = \frac{(2k+1)\pi}{L}$.
We abbreviate the terms in the sums as $H_{ZZ}(k)$ and $H_X(k)$ respectively.
The VITA then amounts to the decoupled expression
\begin{align*}
    \ket{\psi_P(\balpha, \bbeta)} & =
    \bigotimes_{k=0}^{(L-1)/2} \prod_{p=P}^{1} e^{-\beta_{p} H_{X}(k)} e^{-\alpha_{p} H_{ZZ}(k)} \ket{0_k 0_{-k}} \, .
\end{align*}

Since the true ground state of the TFIM Hamiltonian is known, we can compute the fidelity directly. The exact ground state is given by
\begin{equation}
    \label{eq:exact}
    \ket{\psi_{\rm exact}} = \bigotimes_{k=1}^{N/2} \left( u_k + v_k \cdag_k \cdag_{-k}\right) \ket{0_k 0_{-k}}\, .
\end{equation}
with
\begin{align*}
    u_k     & = \sqrt{\frac{E_k + \zeta_k}{2E_k}}, \qquad v_k = i \sqrt{\frac{E_k - \zeta_k}{2E_k}}, \\
    E_k     & = \sqrt{J^2 + h^2 + 2Jh\cos \theta_k},                                                 \\
    \zeta_k & = -h - J\cos\theta_k \,.
\end{align*}

\subsection{Entanglement Entropy}\label{app:Entanglement}

Because the TFIM is dual to a free system, all correlations,
and hence the entanglement entropy (EE),
can be determined from the two-point functions by Wick's theorem
\cite{peschelCalculationReducedDensity2003,
vidalEntanglementQuantumCritical2003,
*latorreGroundStateEntanglement2004}.
It is convenient to introduce Marjorana fermions $a_{2n} = i(c_n - c_n^{\dag})$, $a_{2n-1}=c_n + c_n^{\dag}$ with the $2L \times 2L$ correlation matrix $\expval{a_n a_m} = M_{nm} = \delta_{nm} + i \Gamma_{nm}$
given by
\begin{equation}
    \Gamma_{ij} = \mqty(
    \Pi_0 & \Pi_1 & \cdots & \Pi_{L-1}\\
    \Pi_{-1} & \Pi_0 &  & \vdots\\
    \vdots & &\ddots & \vdots \\
    \Pi_{1-L} & \cdots & \cdots & \Pi_{0}
    )\,,
    \qquad
    \Pi_n = \mqty(
    0 & g_n  \\
    -g_{-n} & 0
    )
\end{equation}
with
\begin{equation}
    \label{eq:G}
    g_n = \ev{a_n a_0} = \ev**{\cdag_n \cdag_0} + \ev**{\cdag_n c_0} - \ev**{c_n c_0} - \ev**{c_n \cdag_0}
\end{equation}

The correlation functions can be found from their Fourier transforms,
\begin{align}
    \label{eq:corr_k}
    \ev*{c_k c_{-k}}         & = u_k v_k^* \\
    \ev*{\cdag_k \cdag_{-k}} & = u_k v_k^* \\
    \ev*{\cdag_k c_k}        & = v_k^2     \\
    \ev*{c_k\cdag_{k}}       & = u_k^2
\end{align}
Using these expressions to compute $g_n$, we have
\begin{align}
    \label{eq:G_uv}
    g_n & = \frac{2}{L} \sum_{k=0}^{(L-1) / 2} \left( 2 \tilde u_k \tilde v_k \sin n\theta_k + \left(\tilde u_k^2 -\tilde v_k^2\right) \cos n\theta_k \right) 
\end{align}
where $\tilde u_k, \tilde v_k$ are real numbers that characterize the state. 
For the exact ground state, the above formula simplifies to 
\begin{align}
    g_n & =  \frac{2}{L} \sum_{k=0}^{(L-1) / 2} \frac{ h \cos n\theta_k + J \cos(n+1)\theta_k}{E_k} \\
\end{align}

The entanglement entropy of a region $A$ can be obtained by restricting the correlation matrix $\Gamma$ to a region $A$ of size $L_A$.
The eigenvalues of the reduced $\Gamma_A$, denoted with $\nu_m$, contribute equally to the entanglement entropy through
\begin{align*}
    \label{eq:entropy}
    S(L_A) & = -\sum_{\nu}\left(\frac{1+\nu_m}{2}\right) \log_2 \left(\frac{1+\nu_m}{2}\right) \\
           & \phantom{=} \, \,
    + \sum_\nu
    \left(\frac{1-\nu_m}{2}\right) \log_2 \left(\frac{1-\nu_m}{2}\right)
\end{align*}
This method scales polynomially in the system size $L$, allowing the study of systems with hundreds of spins.

\begin{figure}[hbtp]
    \centering
    \includegraphics[width=0.9\columnwidth]{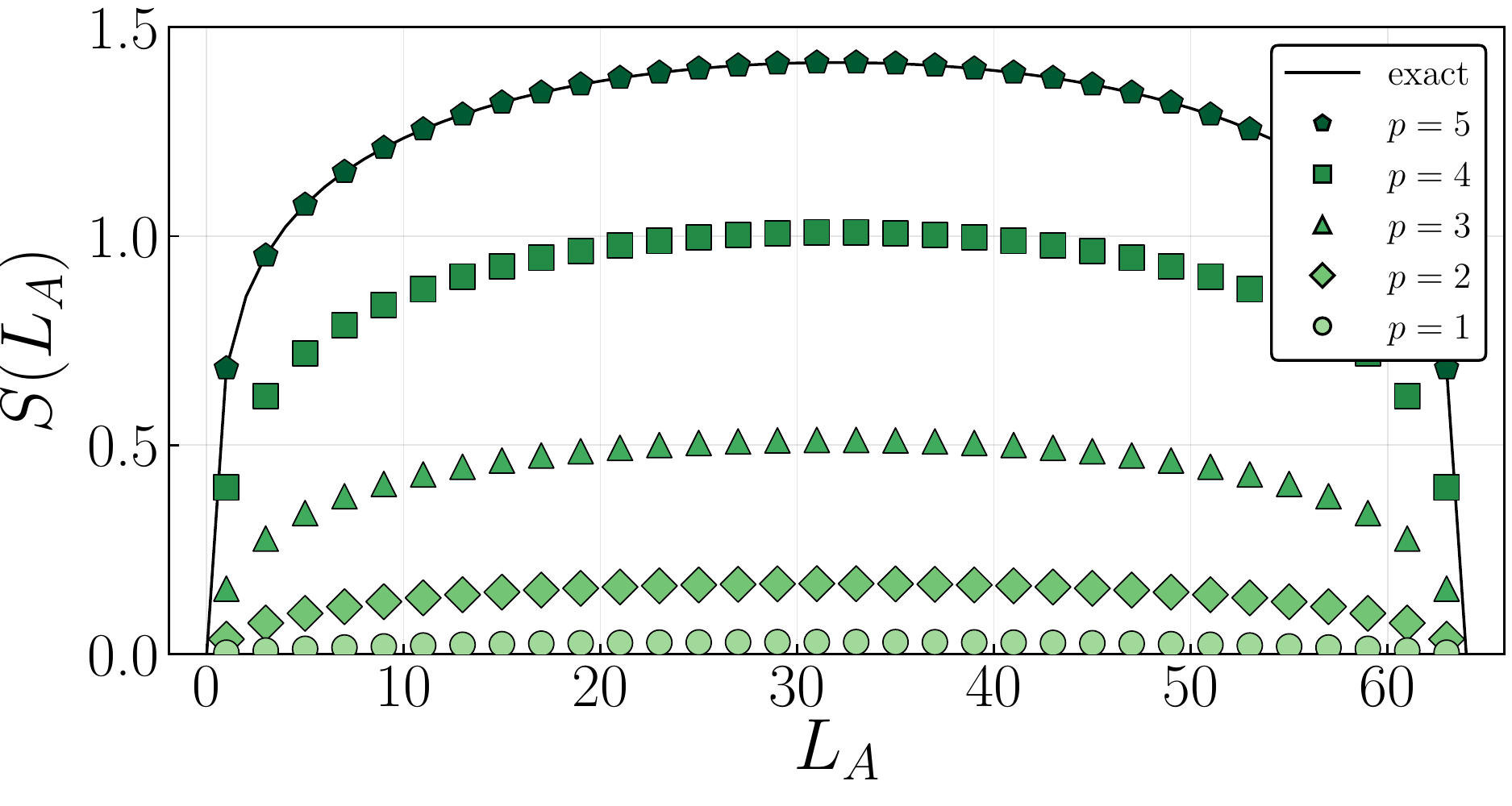}
    \caption{
        \label{fig:ent}
        Entanglement entropy as a function of subsystem size $L_A$ for intermediate $p$ states in a depth $P=5$ ansatz for $L=64$ spins.}
\end{figure}

Figure~\ref{fig:ent} shows the EE of our ansatz converges to the Cardy formula \cite{calabreseEvolutionEntanglementEntropy2005} as $P$ increases .

The spin-spin correlation functions can also be computed from $g_n$.
\begin{align*}
    \expval{X_0 X_n} & = g_0^2 - g_n g_{-n}            \\
    \expval{Z_0 Z_n} & = \det \mqty(
        g_{-1}  & g_{-2}  & \dots  & g_{-n}  \\
        g_0   & g_{-1}  & \dots  & g_{1-n} \\
        \vdots & \vdots & \ddots & \vdots \\
        g_{n-2} & g_{n-3} & \dots  & g_{-1}
    )
\end{align*}

\begin{figure*}[bthp]
    \centering
    \includegraphics[width=1.7\columnwidth]{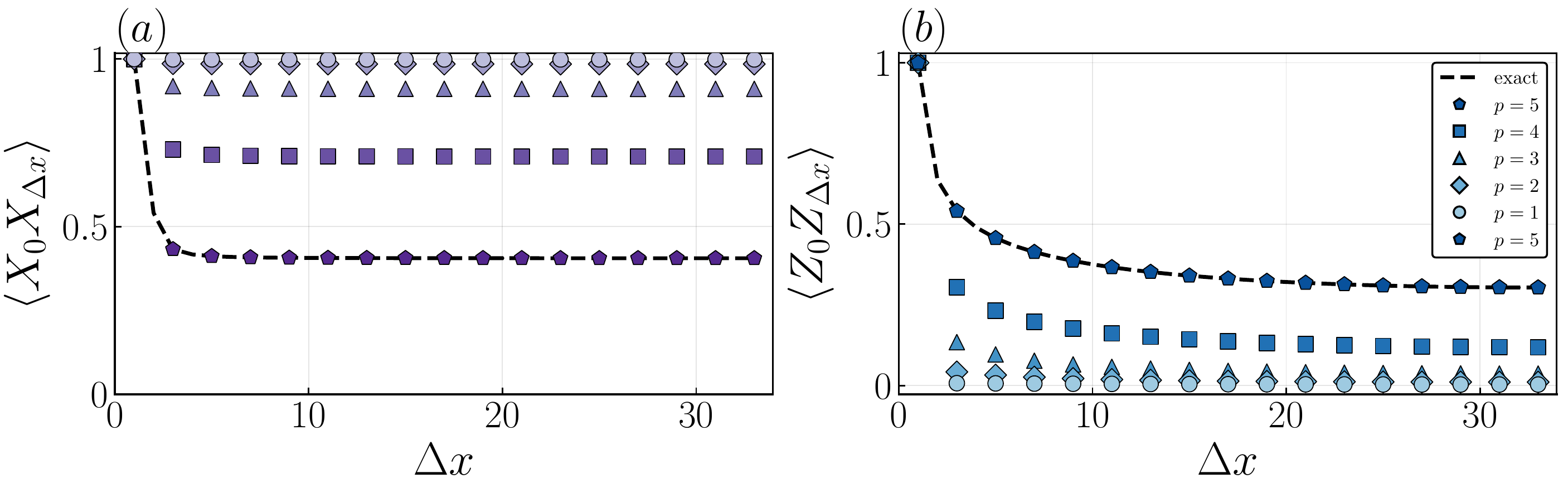}
    \caption{
        \label{fig:corr}
        Spin-spin correlation functions for intermediate $p$ states in a depth $P=5$ ansatz for $L=64$ spins.}
\end{figure*}

Figure~\ref{fig:corr} shows the spin-spin correlations for $P=5$ at the critical point $h/J=1$.

\section{Absence of Light-Cone in Imaginary time evolution}

Here we provide an analytical demonstration that the light cone is not observed in the imaginary time evolution of a pure state. Consider the set-up of a local quench in a $(1+1)$-dimensional CFT described in Ref.~\cite{calabreseEntanglementCorrelationFunctions2007}.  For $ t < 0$, the system consists of two semi-infinite disjoint regions with identical Hamiltonians which are in their respective ground states. At $ t = 0$, the two regions are joined at a point which we choose as the origin. We are interested in the time-evolution of the entanglement entropy for the subregion $ - \infty < x < -\ell$ with the rest of the subsystem (i.e., $ - \ell < x < \infty$). As shown in Ref.~\cite{calabreseEntanglementCorrelationFunctions2007}, for real time evolution, the entanglement entropy takes the following form:

\begin{equation}
    S =
    \begin{cases}
        \frac{c}{6} \log\left(2 \ell/a\right) + c_1\, \hspace{1.3cm} t < \ell \\
        \frac{c}{6} \log\left(\frac{t^2 - \ell^2}{a^2}\right) + c_2\,  \hspace{1cm} t > \ell
    \end{cases}
\end{equation}
where $a$ is the short-distance cut-off and $c_1, c_2$ are constants. The qualitative difference between $ t < \ell$ and $ t > \ell$ originates precisely due to the locality of interactions, which results in a sharp light-cone in the continuum field theory, and a Lieb-Robinson light cone on a lattice-regularized CFT. Heuristically, for $t < \ell$, the quasi-particles generated at location $x = -\ell$ haven't had enough time to reach the entangling boundary.

Next we consider the same set-up except that one evolves the system with imaginary time, which we denote as $\tau$. Mathematically, if we denote the Hamiltonians for $x < 0$ and $x > 0$ as $H_1, H_2$, and the Hamiltonian that connects the two regions at the origin as $H_{12}$, the unnormalized wavefunction for $ \tau > 0$ is given by $|\psi(\tau)\rangle = e^{- \tau (H_1 + H_2 + H_{12})} |\psi(0)\rangle$ where $|\psi(0)\rangle$ is the ground state of $H_1 + H_2$. One can obtain the time-dependence of entanglement entropy following the technique identical to Ref.~\cite{calabreseEntanglementCorrelationFunctions2007}. One finds:

\begin{equation}
    S = \frac{c}{6} \log \left( \frac{2 \sqrt{\ell^2 + \tau^2}}{a}\right) + c_1
\end{equation}
Therefore, although both the real time and the imaginary time entanglement grows logarithmically for times much larger than $\ell$, there is no signature of light-cone in the imaginary time growth of entanglement, and the entanglement entropy starts to increase  instantaneously following the quench.

For completeness, we also discuss imaginary time entanglement growth under a global quench. We now consider the set-up of Ref.~\cite{calabreseEvolutionEntanglementEntropy2005},
where the initial state $|\psi_0\rangle$ is taken to be short-ranged entangled and corresponds to a translationally invariant conformal boundary condition. It was shown in Ref.~\cite{calabreseEvolutionEntanglementEntropy2005} that for \textit{real} time evolution of the state $|\psi_0\rangle$ with the CFT Hamiltonian, the entanglement of a region of size $\ell$ grows as:

\begin{equation}
    S \sim
    \begin{cases}
        \frac{\pi c t }{6a}, \hspace{1cm} t < \ell/2 \\
        \frac{\pi c \ell }{12 a}, \hspace{1cm} t > \ell/2
    \end{cases}
\end{equation}

Therefore, the entanglement becomes volume-law after time $ t = \ell/2$, which can again be understood from a quasi-particle picture. However, under imaginary time evolution, using technique identical to Ref.~\cite{calabreseEvolutionEntanglementEntropy2005} one instead finds:

\begin{equation}
    S \sim \frac{c}{6} \left[\log\left(\frac{2 \tau}{\pi}\right) + \log\left(\frac{e^{\pi \ell/4\tau} - e^{-\pi \ell/4\tau}}{e^{\pi \ell/4\tau} + e^{-\pi \ell/4\tau}}\right) \right]
\end{equation}
Therefore, for $\tau \ll \ell$, $S \sim \frac{c}{3} \log(\tau)$ while, as expected, for $\tau \gg \ell$, $S \sim \frac{c}{3} \log(\ell)$, the ground state entanglement corresponding to the CFT.

\section{Quantum to Classical Map}\label{sec:classicalmap}

The foundation of path integral world-line Monte Carlo is the connection between a quantum system in $d$-dimensions and a classical statistical one in $(d+1)$-dimensions.
In our case, we wish to consider the variational quantum state given by
\begin{align}
    \label{eq:trial3}
    \ket{\psi_P(\balpha,\bbeta)} & = \mathcal{N} \prod_{p=P}^{1}
    e^{-\beta_p H_X} e^{-\alpha_p H_{ZZ}} \ket{+}
\end{align}
where we include the $H_Z$ term for generality.

In the following, we will only consider $P=1$, but the higher-$P$ formulation is nearly identical.
The expectation value of some operator $\mathcal O$, up to a normalization, is
\begin{align}
    \label{eq:operator_map}
    \ev{\mathcal O}
     & = \expval{\,\mathcal O\,}{\psiV}
    \nonumber                                                                                    \\
     & = \mel{+}{\qaoaT{1} \mathcal O \qaoa{1}}{+}
    \nonumber                                                                                    \\
     & = \sum_{\{s\}}\sum_{i=1}^{L} \sum_{p=1}^{2P+1} \tilde{\mathcal{O}}(s_{i,p}) \, p(s_{i,p})
\end{align}

The probability $p(s_{i,p})$ is the Boltzmann factor, given by the exponential of the Hamiltonian
\begin{align}
    H_{\rm 2D} & = -\sum_{i=1}^{L} \sum_{p=1}^{2P+1} \left( J_x(p) s_{i,p} s_{i+1,p} + J_t(p) s_{i,p} s_{i,p+1} \right)
\end{align}
up to a normalization $Z=\sum e^{-H_{\rm 2D}}$,
where the couplings of the classical model are related to $(\balpha,\bbeta)$ via
\begin{align}
    J_x(p) & = \alpha_p                      \\
    J_t(p) & = \frac{1}{2} \ln \coth \beta_p
\end{align}

\begin{figure}[hbtp]
    \centering
    \includegraphics[width=0.95\columnwidth]{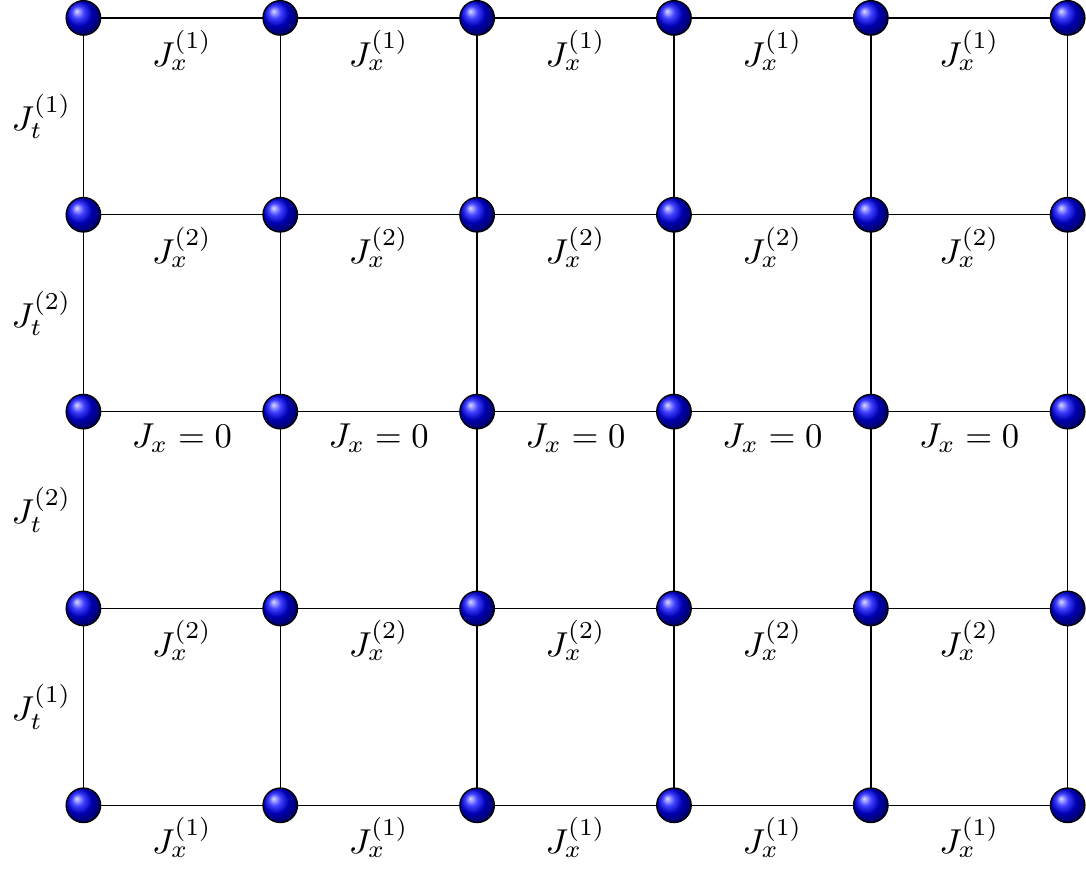}
    \caption{Schematic of the two-dimensional classical Ising lattice dual to the ansatz in Eq.~(\ref{eq:trial3}) for depth $P=2$ and $L=6$ spins.}
    \label{fig:lattice}
\end{figure}

Continuing this process for higher $P$ is straightforward, and can be visualized as a $L\times(2P+1)$ lattice in Fig.~\ref{fig:lattice}.
Adding connections $J_x(0)$ between the middles time-slices ($p = P+1$) one obtains the Suzuki-Trotter inspired neural networks in Ref.~\cite{freitasNeuralNetworkOperations2018}.

The only observables required to compute the total energy are:
\begin{align}
    \ev{Z_i Z_{i+1}} & = \sum_{\{\vb s\}} s_{i, m} \,s_{i, m} \,p(s_{i, p})                  \\
    \ev{X_i}
                     & = \sum_{\{\vb s\}} \exp({-2J_t{(m)}\,s_{i, m}, s_{i, m}}) p(s_{i, p})
\end{align}
where $m=P+1$ denotes the middle time-slice.

The gradients of these expressions can be evaluated via
\begin{align}
    \partial_{\theta} \ev{H} & = \ev{\partial_{\theta} H} - \ev{H \partial_{\theta} \vE} + \expval{\vE}\expval{H}
\end{align}
where $\vE$ is the energy of the classical Ising model, and $\btheta$ denotes $(\balpha, \bbeta)$.
The exponential form of Eq.~(\ref{eq:trial3}) make stochastic reconfiguration (SR) \cite{beccaQuantumMonteCarlo2017}, easy to implement.
SR uses the positive-definite covariance matrix $S$ as a pre-conditioner for the gradient where $S$ is defined as
\begin{align*}
    S_{\theta, \theta'} & =  \ev{\mathcal O_\theta \mathcal O_{\theta'}} - \ev{\mathcal O_\theta} \ev{\mathcal O_{\theta'}}\,,
\end{align*}
with $\mathcal O_\theta= \partial_\theta (\log \psi(\btheta))$.

Each gradient update then takes the form
\begin{equation}
    \theta\leftarrow\theta-\eta S^{-1} \partial_{\theta}\ev{H}
\end{equation}
for a small learning rate $\eta$ which we take to decay during iteration.
It is possible that $S$ is non-invertible so we use the Moore-Penrose pseudo-inverse.

\begin{figure*}[hbtp]
    \centering
    \includegraphics[width=\textwidth]{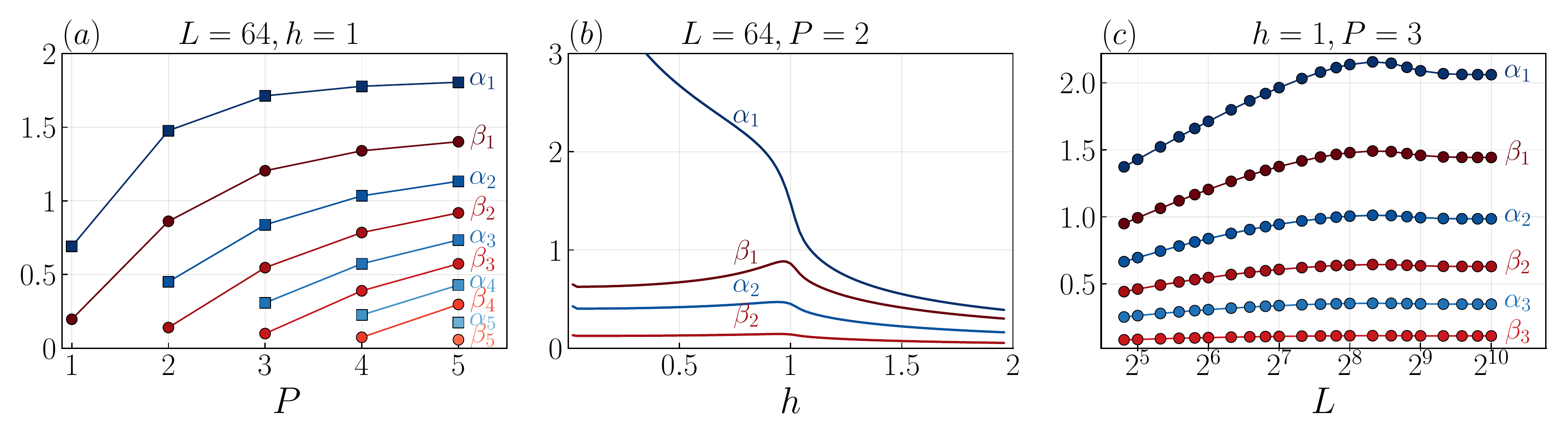}
    \caption{\label{fig:parameters}
        Optimal parameters $(\balpha, \bbeta)$ found using the Jordan-Wigner method for $L=64$ spins for (a) the critical point $h=1$ with $P=1,...,5$, (b) Optimal parameters for $P=2$ for various $h$, (c) scaling of optimal parameters with $L$ at $h=1$.
        All plots are generic for any $P$ and system size $L$.}
\end{figure*}

We typically find optimizing $P=1$ converges rapidly, while higher-$P$ becomes more difficult.
Optimal parameters for a fixed $P$ vary smoothly with system size and field strength, which makes inferring `nearly' optimal parameters easy (Fig.~\ref{fig:parameters}).
This result is highly similar to the patterns in optimal parameters found in Ref.~\cite{zhouQuantumApproximateOptimization2018} for QAOA.

\section{Imaginary Time Required to Prepare GHZ State from Paramagnet}

Consider the case of preparing the GHZ state 
$|\textrm{GHZ}\rangle$,
from the paramagnet $\ket{+}$ using the projection in imaginary time, $e^{-\tau H_{ZZ}}$.
We will show that the total time $\tau$ required scales with $\log L$. 

Suppose that after time $\tau$, the infidelity between the target and projected state is  $\epsilon$,
\begin{equation}
    \label{eq:ghz}
    |\langle {\rm GHZ}|e^{-\tau H_{ZZ}}|+\rangle|^2=1-\epsilon\,.
\end{equation}
Setting the energy of all up spins (or all down) to zero, 
the squared norm of the projected state,
$e^{-\tau H_{ZZ}}|+\rangle$,
 is the partition function of the 1d Ising model at inverse temperature
$2\tau$:
$${\cal Z} = (1-1/x)^L + (1+1/x)^L\,,$$
where $x\equiv e^{4\tau}$.

Taking the limit $L\rightarrow \infty$ with $x=\alpha L$, we find that $ {\cal Z} = 2 \cosh(1/\alpha)$. Therefore we satisfy Eq.~(\ref{eq:ghz})
if 
$\alpha^{-1} = \cosh^{-1} ((1-\epsilon)^{-1})$, 
and 
$\tau =\log(\alpha L)/4$.

\section{Error in Fidelity Provides a Bound on Error in Observables}

Let $\psi_t$ denote the target state and consider a wavefunction $\psi$ such that $1-|\langle \psi_t|\psi\rangle|^2<\epsilon$.  Then there exists $\psi_t^\perp$ such that $\psi = \sqrt{1-\epsilon} \psi_t + \sqrt{\epsilon} \psi_t^\perp$.  Hence, for any observable $O$,
\begin{align*}
    \langle \psi | O|\psi\rangle = & (1-\epsilon) \langle \psi_t | O|\psi_t\rangle \\+&2\sqrt{\epsilon(1-\epsilon)} Re\big(\langle \psi_t^\perp|O|\psi_t\rangle\big)
    + \epsilon \langle \psi_t^\perp | O|\psi_t^\perp\rangle
\end{align*}
If the operator norm of $O$ is $c\equiv \mathrm{sup}_{|v\rangle \neq 0} \frac{||O|v\rangle||}{|||v\rangle||}$, then
\begin{align*}
    \left|\langle \psi | O |\psi\rangle - \langle \psi_t | O|\psi_t\rangle\right|
    \leq 2c \left(  \sqrt{\epsilon(1-\epsilon)} + \epsilon \right).
\end{align*}

\bibliography{paper}